\documentclass[11pt]{article}
\usepackage{geometry}                % See geometry.pdf to learn the layout options. There are lots.
\usepackage{graphicx}
\usepackage{amssymb}
\usepackage{apacite}
\usepackage{authblk}

\usepackage{epstopdf}
\begin{document}

\title{When Science is a Game}
\author{Simon DeDeo}
\affil{{\small Department of Social and Decision Sciences, Carnegie Mellon University, Pittsburgh, PA 15213 \\ Santa Fe Institute, Santa Fe, NM 87501}}
\date{}

\maketitle

\abstract{\noindent What happens when scientists are, at certain points in a field's development, playing a game? I present a framework for such an analysis that draws on the theory of games provided by the historian Johan Huizinga. Huizinga gives five conditions for a social practice to become a game: free engagement, disconnection, boundedness in time and arena, the order-creation of rules, and the presence of tension. Application of this theory to scientific practice predicts patterns of behavior that can be tested by quantitative analysis: the emergence of hard boundaries between disciplines, the closure of loopholes in theory creation, resistance to certain innovations in journal publication, and the ways in which scientists fail to prosecute colleagues who engage in questionable research practices.}

\vspace{1cm} \noindent In its study of the social nature of scientific knowledge-formation, the history and philosophy of science has neglected the idea that its object of study can, at certain periods, take the form of participants playing a game. This may make mysterious, at times, crucial features of both the development of science and its current practice. Being alert to the possibility that science can turn into a game may allow us to explain certain puzzling features of its history, and make predictions about its future evolution.

To be clear, when I talk about scientific practices as a game, I do not mean that they are, as an analogy, \emph{like} games at times, or that, as Kuhn famously said of eras of ``normal'' science, that they share some features with games. This is quite different from the proposal made here that, literally, at certain points in time scientists may sometimes do science in the same way we play Chess or Roulette, Ring a Ring o' Rosie or Dungeons and Dragons.\footnote{One must distinguish between playing a game, the concern of this essay, and ``being playful'', a wider notion that only partially overlaps with the former. Science is often seen as playful---think, for example, of the publicized antics of physicist Richard Feynman---but this is neither necessary nor sufficient to make it a case of playing a game. Anything may be done playfully but not all play is playing a game. The confusion is easily made, unfortunately, and not just because the two senses, of play (on the one hand) and of the playful (on the other) share an etymology, but because scientific creativity itself shares many of the features of the explorations of children \cite{gopnik1996scientist}, a demographic that also spends a lot of time playing games.}

%In popular culture, scientists are commonly depicted as playful children. Popular images of Einstein show him riding a bicycle or sticking out his tongue. Less approvingly, in his report on the Harvard Grant study, George Valiant depicts the absorption of ``William Mitty'' into the study of astronomy as a neurotic prolongation of childhood~\cite{vaillant2012adaptation}. Sigmund Freud's study of Leonardo da Vinci credits his genius with the early, and complete, sublimation of childhood sexual curiosity into research, with the side-effect of preventing development into a psychosexually mature adult~\cite{freud1964leonardo}. We must carefully distinguish the playful attitudes and emotions characteristic of a child, who also sometimes plays games, from the phenomenon of game play itself.}

Understanding this more substantive thesis requires we be clear on an underlying type--token distinction: ``Chess'' is a game that we play, an instance of which involves two players sitting down over a board, seeing that board, and each other, in a certain way, and moving the pieces around. In a similar fashion, this essay will address the possibility that a science can become, at points, a game where the individual instances of game-play are writing, submitting, and publishing a scientific paper. Fields as different as social psychology and theoretical physics may, because of their intellectual content and the social conditions of the field, become games based around the peer-review system. 

Such a game is not against any opponent: the author need not see himself as winning against a reviewer the way one tennis player wins against somone on the other side of the net. The referee might, rather, appear more like a Dungeonmaster in D\&D, or even Captain Hook in a game of Peter Pan. These games may be played alongside other forms of science: one might think to take a journal's table of contents and ask which of the articles therein was the outcome of a game, rather than (for example) an open-ended Popperian or Peircean knowledge-seeking activity.%\footnote{As a practicing scientist, I consider the development of a science into a game a bad thing, in part because of how a (Huizingan) game is, fundamentally, bounded; see further discussion in condition three, below.}

To bring rigor to such an analysis requires drawing a reasonably strict boundary between game and ``like a game''. This, in turn, gives us a demarcation problem parallel to the original one of separating science from pseudoscience. The demarcation problem for game play is easier to solve in part because play is a much older practice than science, is a common activity for much of the population, and is learned earlier in an individual's development. This means that we are reasonably fluent in drawing the distinction and achieving intersubjective agreement about the majority of cases on either side of the game--not-game boundary. The distinction is more than conventional because games are a natural kind in the social sciences: anthropologists see game play in a variety of cultures (see, for example, \citeA{geertz2005deep}'s analysis of the ``deep play'' of the Balinese cockfight), and can observe cross-cultural regularities in the causal consequences of drawing the game--not-game distinction~\cite{games_anthro}.

Once a practice does become a game, new constraints appear that influence its development over time. Take, for example, the game of Chess. Some of the things that happen in Chess (pawns never move backwards, bishops only diagonally) appear permanently fixed, while other patterns, such as those in opening gambits or endgames, undergo high rates of selection and variation. Unless we recognize the characteristic features that make Chess a game rather than, for example, an aesthetic practice of wooden-piece ballet, it is difficult to understand why some things vary and others remain fixed. An identical challenge appears in the case of science, where we struggle to identify the moving parts of a scientific practice and the different time-scales of change. If science is a game at some point, then we might expect to see characteristic timescales of change for different features. We may also see different patterns in its evolution: the game of football springs from rugby, but in a discrete fashion---there is no such thing as an ``interdisciplinary'' football-rugby game, for example.

Seeing science as a form of game play is helpful also because science, like games, often has an unusual relationship to more ordinary forms of life and work. If we do not understand the autonomy of the game from the power structures of the society in which it is embedded, and from the immediate material needs of its participants, we will misunderstand the role of scientific institutions and the ways in which they can, and can not, affect the development of scientific fields. The concept of game play can help us explain the ambiguous freedom of a scientist, and allow us to go beyond a ``skeptical'' description of scientific activity in terms of the shifting power relationships between, say, researchers and grants officers, or, equally, an account that explains the social aspect of science as driven by the difficulties of gaining causal control over nature.

%% TKTKTK
Finally, the game perspective is useful when our interest in science goes beyond the purely scientific. We want to know not just how the replication crisis could have come about, for example, but how to recover from it. In such instances we look at science as a social engineer or activist, because we want to intervene to fix what seems broken. The hypotheses that come from the game perspective may enable us to better predict what kinds of interventions will be more likely to take hold, and which particular interventions are more likely to fix the problem without undesirable side-effects.

Because the games that people play emerge in a social context, the claim that science can be a form of game play looks a lot like the widely-accepted fact that social factors play a role in knowledge formation. This equivalence relies on too weak a notion of what a game is, however, because whether or not scientists are doing science is not just a matter of social convention. If game play is a natural kind we can be wrong about it in ways we can not be wrong about, say, whether or not a scientist has won a prize. We might, for example, agree as a society that certain socially-privileged players in a Chess tournament may move their rooks diagonally, and we might even agree to call the actions executed by these persons ``Chess'', but that agreement does not mean that the people involved are actually playing Chess, or indeed, any kind of game at all.

To establish the connection between science and games requires both that we have a clear account of the constitutive characteristics of a game, and that we match these characteristics against either what we see happening in science, or what we can see in the normative accounts of what scientists believe they are doing. While I will adhere closely to a particular interpretation of game play due to Johan Huizinga, my descriptions of scientific practice will be non-dogmatic, informal, and eclectic. This has the advantage of providing a systematic introduction to the concept of game play for historians and philosophers of science, while at the same time allowing readers to make connections to a number of different interpretive traditions.

\section{The Huizingian Theory of Play}

Play---meaning, in particular, the playing of games---is an understudied phenomenon. When it draws the attention of science it is often for its epiphenomena: play as a tool for learning, as a social force in a ritual context, or as a test case in the study of algorithms or game theory. In my own investigations, I find two traditions that attempt to study game play on its own terms. The first, represented by projects such as the MIT Press Game Design series, is the contemporary field of video game studies: just as literary theorists and departments of creative writing must study plot and rhetoric, so must video game designers engage with what audiences do with an object meant to be played as a game.

This contemporary tradition of game design is young, but it connects to an older, humanistic tradition that begins with Johan Huizinga's 1938 book \emph{Homo Ludens}~\cite{huizinga2016homo}. Huizinga is the first modern scholar to develop a theory of play that clearly distinguishes it from nearby human activities such as artistic creation and contractual agreement. Notably, for Huizinga the cognitive aspects of play, a preoccupation of accounts of play in child development or animal behavior, are secondary. So, too, are play's sociological aspects; indeed, Huizinga will locate the origin of key features of social life in game play, rather than the other way around.

Huizinga enumerates five necessary characteristics associated with play that will be relevant to us. {\bf First}, play is engaged in \emph{freely}, \emph{i.e.}, there is no moral duty or material urgency to play, nor can play be demanded by another person. If it is, it ceases to be play. A player may feel compelled to play for reasons of pride or even instinct, of course. Such compulsions have a similar relationship to play that they do to other necessarily un-coerced actions. Play shares the necessity of freedom, for example, with making an apology, and compulsion vitiates its force.

The norm of free engagement is also characteristic of scientific practice, and not just because of the belief that professional work is a vocation rather than a matter of practical necessity. It is also contained in the specific separation of theoretical science from the practical demands of engineering. Under a strict version of these norms, scientists may address practical problems without violating the free-engagement criterion, but only in as much as they consider this a side-benefit rather than a material motivation. A scientist may turn to practical endeavors, but (under this norm) on the occasions she does so, she is not acting as a scientist. The idea of free engagement clashes, notably, with the common-task cooperation between the theoretical and applied sciences in Francis Bacon's \emph{New Atlantis}: on Bacon's island, the task of the scientist is defined primarily in terms of his duties to the wider society.

{\bf Second}, play involves a \emph{disconnection} from reality: the King in Chess is not a monarch, nor is Captain Hook in a children's game of Peter Pan a real captain. In order to play a game, the player must pretend to inhabit a world where the rules and meanings of ordinary life do not apply; in Huizinga's anecdote, the child playing at being a train tells his father ``don't kiss the engine, Daddy, or the carriages won't think it's real''. 

The disconnection characteristic of game play often goes unnoticed. Consider the car-trip game of license plates, where one counts license plates with the goal of, for example, being to be the first to ``catch'' all fifty States. The arrival of the plate is, like the dice in roulette, subject to a near infinite number of random real world events, but what matters is a single surface feature (listed state) and not (for example) the functional role it plays in the larger scheme and the utility it has, for example, to a police officer on patrol or an insurance agent adjusting claims.

On a larger scale, consider the Fischer-Spassky match of 1972, when an American and a Soviet Chess grandmaster faced each other in a series of games. What is compelling  about this encounter is precisely the extent to which the rules of the Cold War were suspended and transposed: the relative positions of the two superpowers are irrelevant to a game defined solely in terms of the pieces on the board. The most extreme case of the disconnection games enable is the Christmas day football match that took place on No Man's Land during the trench warfare of World War One. This strange game was played by members of the Royal Warwickshire Regiment and an enemy unit of the German Army who just prior had been engaged in deadly warfare. The game was not the cause of this unusual ``Christmas truce''; rather, the emergence of an (implicit) truce was necessary for it to be a game at all. Indeed, in this case one might say that in playing the game, the truce that obtained was completely distinct in character from the truce that would, one day, end the war. It was a consequence of game disconnection, rather than contractual agreement. From the point of view of the nations, war resumed the next day as if nothing had happened. 

The pretense has both social and epistemic consequences. Socially, ``just pretend'' gives play a disinterestedness that other activities lack. A player who loses to his boss for expediency's sake, or to end the match in time for tea, is no longer playing the game. The real has intruded and broken a spell. The social disconnection of ``the scientist'' is a familiar fable, found in the legend of the death of Archimedes and the Hollywood meme of the absent-minded professor. Less remarked upon, however, is the disconnection of the epistemic objects of science itself: it is not just that the child can't be kissed by his father because he ``is'' a train, but also that the child knowingly coexists as both child and train. The Queen in a pack of cards is no queen; but in certain forms of science neither is the variable in a model a variable of the world. 

Much of science has this kind of as-if structure~\cite{appiah2017if} because the components of a scientific framework are not (usually) taken to be in themselves elements of the world. Process models in cognitive science, statistical models of epistatic interactions in biology, and the ``effective'' field theories of quantum theory all work with objects whose properties and meanings are partly autonomous from the events that occur in the laboratory or field site. The neurological instantiation of a mental state, for example, is not part of the model for a psychologist who creates a model based on mental states to explain an experimental result. The mental states of that  model may be equally invisible to introspection and to an fMRI scan without affecting the validity of the psychologist's game. This as-if structure is preserved by a parallel agreement on what counts as experimental evidence. A microeconomic model of rational actors is not undermined by the results of an experiment in behavioral economics---unless, of course, the modelers agree on this ahead of time. 

What is ruled out is just as important as what is ruled in, because what is ruled out generally cannot count as evidence, even in a positive sense. An economist presenting a utility-theoretic model of human behavior, for example, may appeal to a psychological or sociological account of desire but this is taken as simply setting the stage, rather than constituting a part of her economic science proper. ``Need not'' becomes equivalent to ``ought not''; if she appeals to such out of bound considerations too often, suspicions arise as to whether she is still contributing to the science of microeconomics at all. Conversely, an economist presented with a rival's demand to explain something out of bounds has the right to complain that this is unfair. When we see such instances of policing over what forms of evidence count in a scientific debate, we may take them as signals that a game is in progress.

Once the boundaries of a model have been drawn, the experimental evidence that counts necessarily takes on an unreal aspect. Agents of the real world (in contrast to those of a model) are simultaneously biological, psychological, and economic creatures. Any branch of the social sciences, however, must in the course of an investigation  deny some of these facts. A study of the endowment effect, for example, will control for gender before comparing its results to the standard model.\footnote{The endowment effect: ``people show the tendency to demand much more in order to give up an object than they are willing to pay to acquire it''~\cite{roeckelein2006elsevier}.} This is not because gender is not an important part of a subject's response to the endowment, but precisely because it is. An agent's response to being given a mug lies at the center of a enormous number of causes that any Baconian science (or, indeed, commercially-oriented market campaign) would need to account for, but the majority of these effects lie outside of the endowment effect model's scope.

The rules of the game, in other words, permit the scientist to take the objects in abstraction from their causal properties in the world itself. The idea of a ``stylized fact'', captures by means of an artistic metaphor how play objects disconnect from their referents: a King in Chess is a stylized representation of a certain kind of power just as a desire in belief-desire-action theory is a stylized representation of the desires that form the basis of lived experience. ``Operationalization'' is another method of disconnection that turns a causally-complex phenomenon (\emph{e.g.}, economic prosperity) into a potential game object (\emph{e.g.}, GDP per capita).

Disconnection means that models may even predict their own impossibility, \emph{i.e.}, their truth can be incompatible with the very fact of their being able to be written down. A Bayesian model of cognition may be developed even if the assumptions of the model imply that the preconditions of the mind it describes arose from a non-Bayesian process; not only that, but the process by which the model itself is developed---\emph{i.e.}, the reasoning process of the scientist who creates it---may also be incompatible (she may, for example, recognize that she never had the possibility of such a model in her space of priors before she began thinking about it). A libertarian theorist of free-market economics may accept a tenured position at a public university, and at the same time make a case for private enterprise that implies that his own employment relationship can never produce good science.

Without seeing the pervasive quality of this characteristic disconnection, model-building can seem like a purely epistemic process: scientists build models because models help them know things about the world. Yet there is no reason to believe that model construction itself should provide this kind of knowledge, because model construction limits \emph{a priori} the objects and properties that can play a part in the explanation. If one's goal in model-building is the prediction and control of the natural world, then the model must survive empirical tests that are not restricted to the domain of the model itself. If facts about the Prisoner's Dilemma are to be a form of knowledge about the world, it is not enough to show that individuals will play the game according to this or that Nash Equilibrium; rather, it is that they do so in real-world situations that makes the Prisoner's Dilemma a discovery \emph{about} something. It is, however, just this reality, in its messy and uncontainable form, that game theory attempts to exclude.

Whether or not it comes in the form of an as-if model, the disconnection characteristic limits the set of scientific practices that can, potentially, be considered as games. It is certainly contrary to the understanding of science as an attempt to discover the causes of things. When science uses causal control as a criterion it opens itself up to tests of unlimited scope; any randomized control trial, for example, can fall victim to the influence of a previously unrecognized common cause. In as much as a scientist is engaged in a cause-seeking task, her experience is beyond the bounds of the ``pre-statable''~\cite{stu}.\footnote{That games allow for these rather extreme forms of disconnection has a parallel in the mathematical community, where G\"{o}delian impossibility results lead mathematicians to abandon simple forms of Platonism that make theorems answerable to something outside the game of theorem-proving. The parallel to the Christmas truce or the Fischer-Spassky match is clear: in each case, the rules of the game are incompatible with the preconditions that in fact initiated the game (\emph{e.g.}, a Clauswitzian ``total war'', hot or cold).} Popperian falsifiability is similarly excluded from the domain of playable games because of a similar lack of constraint on what can intervene from outside. Popperian theories, and derivatives such as those presented by \citeA{gelman2013philosophy}, are answerable to effects beyond the scope of the models in which they are expressed.

William Whewell's notion of consilience is another example of how epistemic concerns clash with the disconnection condition. In consilience, scientific value comes from prediction not only out of sample (\emph{i.e.}, from replication), but when an idea ``enables us to explain and determine cases of a kind different from those which were contemplated in the formation of our hypothesis'', where value stems from the ``remoteness'' of the new cases~\cite{whe}. Formalized in the contemporary language of inference, the explanatory value of consilience is driven by two things: first, that the new cases appear after the presentation of the hypothesis, and second, the conceptual distance to those new cases~\cite{wojtowicz2020probability}. An extended case against the disconnection necessary for scientific activity to become a game is found in the account of knowledge-formation associated with pragmatism and Charles Sanders Peirce in particular. The claim of \citeA{peirce1868some}, that ``the existence of thought now depends on what is to be hereafter; so that it has only a potential existence, dependent on the future thought of the community'', argues not against the pretense of an as-if claim, but the (pre-statable) disconnection of that pretense from the activities of the community as a whole.

{\bf Third}, play is \emph{bounded}: any game has limits of time and place, a beginning and an ending, and a meaning that can be contained within that frame. A game that is over can not be continued, except by a (freely engaged-in) opening of another round. The boundedness of game play has parallels in art: a painting goes up to the edge of a canvas, a piece to the end of the score, a play to the final act, but no more. Huizinga places particular emphasis on boundedness in space: one freely steps into a ``magic circle'', whose geometry delimits the area of disconnection. Even in the internet era, a spatial metaphor persists: a game is confined to a virtual world, a bulletin board, an interface. The boundedness of play, temporal or spatial, means that we know when the game is on, and when it is not. The football match of the Christmas truce must, finally, come to an end.

This criterion of boundedness needs to be distinguished from the way in which effects are ruled in or out of bounds for the purposes of disconnection. Boundedness here is to be interpreted literally because for a scientific activity to count as game play it must have a finite duration in time, and it must take place within an arena. That arena may be partly metaphorical. It is possible to play Chess by mail, over an unspecified geographic extent, yet still be engaged in a game because the enlargement of the geographic space that the players inhabit is not relevant to the game itself, which takes place on a bounded grid eight squares on a side and over a finite duration of time. A crossword puzzle may take minutes, days, or even months to solve, and yet the puzzle is spatially and temporally bounded: when the grid is solved, the game terminates.

The boundedness criterion matters because it tells us how to find instances of the game being played. The most obvious boundary of this kind in contemporary science is the spatial extent of the journal article and the temporal process of peer review that creates it. A journal article is understood as a self-contained object to be judged on what appears within. It is bad form, as a referee, to give a poor article a pass because one knows (say) that the authors really did do the work and simply didn't put it in, or know the caveats that the article does not provide, or because the authors have done much better work elsewhere that they never got around to publishing. Double-blind reviewing, in detaching the object of evaluation from the unbounded context of the author's career, makes this boundedness explicit. 

A referee's judgement encompasses more than the text itself, of course: the referee might judge the quality of the citations to what has come before, or its relevance in the context of the field. These judgements, however, are always restricted to what appears in the text, just as much as the correctness of a crossword solution depends on the questions even when the questions refer to facts outside the puzzle itself. A relevant gap in the reasoning is one that appears within the text itself, while a gap that appears in the reasoning of something the article cites is problematic only because of the effect it has on that in the article: another citation must be found, or a different argument made. A referee that considers that gap irremediable (\emph{e.g.}, because she thinks the entire basis of the subfield is flawed) is expected either to recuse herself, or to make a case that the subfield in question should not be a subject within the journal's remit. In the latter case, she can be expected to be overruled without the journal compromising the rules of play.

Journal publication is similarly bounded in time by the familiar steps in the publication process: received, sent for review, reports received, responses, editor decision, and so forth. Publication is bounded by acceptance or rejection and the exhaustion of appeals. Upon publication, the game is over and cannot be further extended.

Even retraction respects the boundedness of the journal game. A scientist who presents fraudulent data to a colleague before its publication---or even in the process of review---can correct his mistake or (at worst) pull the paper from consideration. A published article that contains fraudulent data, however, can only be retracted as a cheat; the retraction notice plays the same role as the asterisk attached to a sporting record tainted by a doping scandal. Preprint servers such as the Physics ArXiv provide a limit case. On the ArXiv, a preprint may be replaced an arbitrary number of times, and while prior versions are available if desired, readers are directed without remark to the latest version. There is no end to the number of updates on the ArXiv, and thus no game can be played; in this open-endedness the preprint system is closer to the Republic of Letters that preceded the Royal Society, than to the journal article format that institution helped define.

{\bf Fourth}, play has \emph{rules}; in Huizinga's phrasing ``play creates order, \emph{is} order''. Games suspend the rules of ordinary life, but they do so only to impose an even-more binding set. A professor may cut a student some slack by extending a deadline, but a player in Chess can not grant his opponent an extra move just because he is in trouble.\footnote{Or, rather, to be clear: he may do so, but only at the risk of ceasing to play a game; this in the same fashion as the privileged player granted the ability to move his Rook diagonally.} In this, play is stricter than music: a composer may dare to break the rules of his time, and if he does well enough he may be celebrated for it. No such exception exists in the case of play; ``into an imperfect world and into the confusion of life,'' writes Huizinga, play ``brings a temporary, a limited perfection.''

Rules of this form in science are easy to find. Depending on the field one does linear regression (not logistic), controls for gender (but not class), tests for order effects (but not time-of-day), hits a $p$-value of $0.05$ (but not $0.06$)---and the effect is publishable. A literal reading of each of these practices leads to nonsense: negative probabilities, the idea that rich and poor value the same things, that morning is not evening, that one-in-twenty is unlikely but one-in-sixteen is not. In the game of science, however, they demarcate, fairly for all players, what the boundaries are, what is fair play, and what is a fair win. Huizingan rules are the ordering principle of the game itself, not just (or only) constraints or recommendations that appear as part of a larger task.

Perhaps the most famous description of rule-based scientific practice that meets the Huizingian criterion comes from the Kuhnian notion of ``normal science'' found in \emph{The Structure of Scientific Revolutions}. Kuhn's introduction of the term makes the parallel to games such as ``crossword puzzles, riddles, Chess problems, and so on'' explicit; the work of normal science is puzzle solving, constrained by ``rules that limit both the nature of acceptable solutions and the steps by which they are to be obtained''~\cite{kuhn}. (In his account of scientists as puzzle addicts, Kuhn in passing also gestures toward ways of doing science that are \emph {unlike} puzzle solving, and his list includes those scientific practices motivated by ``the desire to be useful, the excitement of exploring new territory, the hope of finding order, and the drive to test established knowledge.'')

One of Kuhn's examples of rule-based scientific puzzle-solving is the suggestion, made in the 18th Century in the light of empirical evidence, that Newton's law of gravity might deviate at small distances from the inverse-square law. For the normal science of the time, this revision was outside the rules of the game; the integrity of the game required the introduction of new celestial bodies to restore the predictive power of the inverse-square law. This kind of rule is common in the physical sciences, which often pose their puzzles in terms of mathematical constraints that it is cheating to defy. A contemporary example is the Lorentz invariance, the relativistic version of Galileo's principles of uniform motion. Certain mathematical inconsistencies in quantum field theory can be eliminated by fiat if one violates Lorentz invariance, which can often be done in a way that is invisible to current experimental tests. The rule that it must be preserved makes these inconsistencies much harder to solve, and is the source of a major ``puzzle-solving'' thread in 20th Century physics known as renormalization~\cite{peskin2018introduction}.

A second example of a mathematical rule for the game of normal science comes from cosmology and astrophysics. There, a rule associated with the dominant community is that one must preserve basic symmetries, found in the Einstein equations, that connect the curvature of space-time to the distribution of matter and energy. The existence of both dark matter and dark energy are consequences of the game built around this rule, while the rule-breaking alternative, known as MOND (Modified Newtonian Dynamics), amounts to a competing paradigm~\cite{mcgaugh2014tale}. While the rules of play in the ``Einstein equation'' game are provided by mathematical constraints, the MOND game is defined much more by the domain of what is to be explained, which, in this case, is the phenomenology of galactic rotation curves. In game conditions, the two paradigms are incommensurable in the way that Go and Chess are incommensurable.

Of course, ``transformative'' science is full of examples of scientists breaking the ``rules of their time''; this may be a purely intellectual matter (renouncing this or that \emph{sine qua non} deductive claim), or something more socially and historically situated~\cite{feyerabend1993against}. Kuhn, to be clear, does not say that its opposite, normal science, is a game in the Huizingan sense (or any other). His goal is rather to draw attention to a certain game-like feature of what happens in certain periods. The order-making nature of rules is necessary for a practice to become a Huizingan game, but not sufficient.

{\bf Fifth}, and (for our purposes) finally, play is characterized by \emph{tension}. Tension, in Huizinga's picture, supplants the more na\"{i}ve idea that in a game players want to win. For Huizinga, not all things we want to call games have a simple win condition (consider, for example, the play that Dungeons and Dragons players engage in), but all have a notion of success that the players find compelling and strive for.

The tension need not be agonistic, in the sense that one player seeks to accomplish the identical thing to another, only better; rather, players experience tension so long as they face the risk of failure in some sense. Tension is a combination of two effects: uncertainty, which leads players to ``strive to decide the issue'', and the harmony provided by the rules of the game, which creates, or allows to emerge, the thing one strives for.\footnote{Recall that, by the disconnection characteristic, whatever ``the issue'' is must be internal to the game itself.} The tension needs to be internal to a particular instantiation of the game, and so, e.g., a scientist's desire to win the Nobel Prize, or gain standing as a Full Professor, can not count under the Huizingan picture.

One obvious source of tension is the desire to win, and if the game is won by publishing an article, then it seems that the problem of tension has been solved. Things are not so simple, however, and not all games are guaranteed tension solely by virtue of clear win conditions. Players might, for example, find escape clauses in the rules that make wins trivial, or in some sense unsatisfying to achieve; we consider this problem further in Sec.~\ref{cgt}.

\section{Four Hypotheses}

Huizinga's criteria suggest that it is indeed possible for science to become a game. When scientists see themselves as freely engaging in the decision to publish, when model-building or operationalization disconnect the content of their submissions from the demands of reality, when the rules of those models are sufficiently well-defined as to create a puzzle-like order, and when participants maintain the tension of the system by creating technical barriers to publication (discussed further below), the bounded nature and centrality of peer-reviewed research publication may be sufficient for a game world to emerge.\footnote{For practicing scientists, the rules and fair play conventions of peer review may be the most salient aspects of a practice that come to mind when they consider the possibility that they are playing a game. At least in the Huizingan framework, however, facts about the peer review process alone are insufficient. Just because scientists ``play by the rules'' in peer review does not mean they are playing a \emph{game}.}

This section presents four hypotheses about the nature of game-like science. Two of these hypotheses (``rigid but cosmopolitan scales'' and ``magic circles resist innovation'') make predictions for future investigations in the science of science. The other two (``players close loopholes to preserve tension'' and ``sportingness hinders punishments for cheating'') provide explanations within the Huizingan framework for aspects of scientific practice that seem to undermine a purely epistemic view of how science works.

\subsection{Hypothesis one: games create rigid but cosmopolitan scales}

In contrast to the informal heuristics and rules of thumb that accompany most human activity, the rules necessary for play are (from within the game) ``perfect''. One is either playing Chess by the rules or not, and only the most minor variations are expected to accumulate over time. Over two thousand years of development, for example, the game of Go is essentially the same in the Chinese, Japanese, and Korean traditions, with the differences that do exist affecting less than one in ten thousand of the games played at competition level~\cite{ko}. Games may be cultural artifacts (Go emerged from somewhere), but once in existence they tend to persist with a self-sufficiency that transcends cultural origins. At the same time, the all-or-nothing character of game rules means that games do not merge; there is no such thing as a hybrid Chess-Go that is playable at once, or even separately, as both Go and as Chess. Boundaries may be cosmopolitan and transcend cultures of origin, but they are hard.

The account of science as a game thus predicts the existence of a characteristic scale that separates scientific communities within a particular field. This scale corresponds to the boundaries between game systems. Within those boundaries, the strictness of the game concept means that even when individuals have differing interests, educational backgrounds, or social ties, they will be able to play together. Across, however, the disconnection is complete, even when scientists are apparently working on the same problem.

Metaphorically, the games that comprise science may be characterized as a series of water-filled compartments, with perfect mixing within, but not across, wall boundaries. We expect sharp boundaries within a field, but we should not expect these boundaries to nest at smaller scales: there is some characteristic scale at which the game is played, rather than a continuum of more or less densely-linked communities. Cosmologists who play the Einstein game in a particular publication will cite freely within that game's tradition, but will not reference the alternative MOND tradition even when the theories bear on the same phenomena. The citation network within the Einstein game will have short path distance: any particular group will have preferential citation patterns, but these communities will overlap so that no article within the tradition is particularly far from any other. This is not due to prejudice against MOND itself: the empirical results that MOND is particularly good at addressing are fair game for non-MOND players, so long as they follow the rules of the Einstein equation game in addressing them, \emph{e.g.}, the ``chameleon gravity'' of \citeA{berezhiani2015theory}. Conversely, two scientists who are socially connected (because, for example, they are in the same department), but involved in different games, will fail to cite each other despite the social impulse (exterior to the game rules) to confer on each other the benefits of citation.

The compartmentalization predicted to emerge when science is a form of game play is different from that we would expect to arise from differing sets of beliefs about (for example) what approaches are more likely to work or what phenomena are more or less important to explain. In comparison to playable games, belief systems have fuzzy boundaries: Philadelphia Quakers and Mid-Western Quakers differ in points of belief, but are able to collaborate in religious observance more easily than they can with members of other Protestant sects, who can collaborate more easily than with groups yet more distant. Thus a single prayer may fulfill the religious requirements of several different believers simultaneously; or they may negotiate theologically in order to arrive at one that does. By contrast, a piece on a board can only be moved in one game at a time; the same move cannot simultaneously advance two different games. The statements of creed in different religions can be subjected to a notion of overlap, whereas games are always equally and completely distinct from each other from the point of view of their rulesets (Chess, Go, and Bridge).

For belief systems, in other words, no hard compartments can be found and there is no scale below which collaboration is frictionless and above which it is impossible. This church and that church may overlap in some ways but not others, and we expect to see smooth shifts in collaboration and cooperation as we move to increasingly distinct forms of belief. Conversely, the unity provided by a belief system is vulnerable to the balkanization of geographic or political boundaries: a system of belief has nuance that a game does not and believers that meet must negotiate to find common ground. The structures that emerge when practices are arranged around belief systems contrast with the binary in--out distinction found in scientific games.

\subsection{Hypothesis two: players close loopholes to preserve tension}
\label{cgt}

Rules have unexpected consequences, and this accounts for the dynamism that might be dismissed by someone who views normal science as static. Some rule consequences are unacceptable: in particular, the discovery, within the ruleset, of any consequence that violates the tension characteristic. Rules may be discovered to imply that, contrary to appearances, anything goes, and the possible set of outcomes is sufficiently broad to nullify the possibility of anything being at stake. When this happens, we expect these violations will be ruled out of bounds by the players by tacit or explicit agreement.

An example of tension maintenance can be found in the history of ``game theory'' (GT).\footnote{To avoid confusion, I will use the abbreviation ``GT'' to refer to the scientific practice associated with the analysis of interactions using payoff matrices and decision theory, introduced by \citeA{morgenstern2020theory}.} GT is a rule-bound game whose constraints, in a similar fashion to many investigations in physics, are given mathematically. Its rules allow for the creation of new descriptions of social interactions and the search for new equilibria, or predicted outcomes. The rules are sufficiently generative to make it possible for players to vie with each other to create, and solve, a variety of compelling puzzles with (disconnected) analogies to observed social dilemmas.

A consequence of these rules, however, are the folk theorems: under very minimal assumptions, GT can predict an arbitrary set of outcomes for the agents in question. From one point of view, these theorems are very good news: GT turns out to be flexible enough that it can show how outcomes decouple from incentives, and reveals that economic behavior must be understood in the context of a larger project of cultural evolution. If GT is to be played as a game, however, this boon becomes a disaster because it eliminates all possible tension. The folk theorems provide simple recipes for constructing solutions that correspond to these outcomes, which means that a player can no longer demonstrate his prowess by replicating a stylized social fact within the rules of the game. In the language of computer games, folk theorems amount to a cheat code that, if used, takes the player out of the game.

For GT to be playable, tension must be restored. In the history of the field, this is accomplished by the introduction of ``equilibrium refinements'': restrictions on the nature of the solutions that restore the challenge of solution finding~\cite{kreps1990game}. Much like the payoff matrices that form the basis of GT's analyses, these refinements have at best a symbolic relationship to features of the real world.

One example of this in GT is the ``Markov perfect'' refinement for the duopoly problem~\cite{maskin1988theory1}: competing sellers of an identical good (\emph{e.g.}, gas stations) are modeled as interacting with each other on a memoryless basis: a gas station owner responds to being undercut by a competitor in a way that neglects any previous price movements~\cite{maskin1988theory}. Such a restriction provides no benefit to the understanding of the social fact of competition and collusion, because we never suspected that business owners suffer from hourly amnesia; nor do the authors make a case that such an amnesiac system might be found in other parts of the economy. The introduction of a Markov-perfect refinement does, however, provide a way for the solution's authors to demonstrate their mathematical skill within the rules of the game.

From the game of equilibria and refinements, now sometimes known as ``classical'' game theory (CGT), has emerged a new game, evolutionary game theory (EGT; see, \emph{e.g.}, ~\citeA{gintis2000game}). While EGT has a superficial resemblance to CGT (most notably in its use of ``Evolutionarily Stable Strategy'', an analog to a one-shot Nash Equilibrium), it has an entirely different set of rules that allows departures from equilibrium, because strategies are dynamically updated by players. In parallel to CGT, EGT's rules are also constraining: in this case, by a restriction on update rules to single-step updates compatible with a non-teleological rule. While CGT, in other words, required equilibrium but cognitively-unlimited players, EGT allows for non-equilibrium conditions but with strong constraints on the foresight and reasoning capacity of the agents.

That both CGT and EGT can be understood as games does not mean that they fail to provide insight into the human condition. In the case of mechanism design, for example, CGT may allow for the construction of better auction systems that avoid the winner's curse as long as the players are bound (by law, for example) to place bids that respect the assumptions of the model itself. EGT can provide parsimonious explanations for phenomena observed in both the human and animal worlds. In both cases, the domains are limited by the extent to which the real world (approximately) enforces the assumptions of the field. The rewards that flow to a player of such games (in terms of journal publication, for example), however, do not depend upon this connection to reality.

\subsection{Hypothesis three: journal ``arenas'' resist innovation}

As we have seen, journal publication is one way for a scientific practice to achieve the temporal and spatial boundaries necessary for a playable game. This implies that innovations that violate these boundaries will encounter resistance that can not be explained through a purely epistemic or sociological view of the scientific process. Conversely, it implies that those who wish to introduce innovations will do so in a way that attempts to respect the rules of the game.

We see this in the community response to $p$-hacking, or the testing of multiple hypotheses until a result is found that meets the minimum standard of publishable significance, and the post-hoc construction of a paper around that result as if it were the only one tested. It may be a major source of the replication crisis, where a large fraction of major studies are now understood to fail a basic test: the experiment can not be replicated by an independent laboratory.

A study of published results found that ``$p$-hacking'' is widespread~\cite{phack} and may be implicit in the incentive structure for publication itself~\cite{smaldino2016natural}. If the studies of the phenomenon are correct, $p$-hacking is a major epistemic crisis because it leads us to believe in effects that are, in fact, simply fluctuations in a world where those effects do not exist. It is a crisis with a well-understood origin that the communities affected have sufficient training to understand. There is also a simple, easily-understood regulatory response: pre-registration of hypotheses. Under the science-as-game perspective, however, pre-registration runs up against the temporal and spatial boundaries of the journal system, and thus encounters resistance.

In the classical publication game that begins with the submission of an article to a journal, a referee can reject a paper because she disagrees with the quality of the experiment or the nature of the analysis, and an author can respond by conducting more experiments or re-doing an analysis, and thus continue playing. Pre-registration, by contrast, pre-commits the player to a particular subset of moves that she can make in the article itself---there is no such thing as a post-submission re-pre-registration---and thus the ``win" condition, the validity of the publication, becomes dependent on a move made before publication. Somewhat like a Chess player agreeing to an opening ahead of time, the game has begun early.

Open data and reproducibility requirements also violate the boundaries of journal publication, in this case by altering the spatial extent of the game: an article no longer stands alone, but depends for its validity on a set of objects beyond the magic circle of the article itself.

The Huizingian point of view suggests the route that a successful response to the $p$-hacking crisis might take. Pre-registration shifts the temporal boundaries of the field of play, but does not have to eliminate them. A natural solution is to incorporate the pre-registration stage into the publication game itself, rather than having it be a criterion engaged before the game begins. A journal that begins the publication process with a logged pre-registration might find more willing game-players. An open data requirement can be brought within the magic circle by a system of within-article accreditation, an idea that can be found in the Open Science Foundation's system of ``badges'' that accompany articles and serve as a validation, residing within the confines of the article and thus the scope of game play, that the conditions have been met.

\subsection{Hypothesis four: sportingness hinders punishments for bad behavior}

Fairness is a broad concept in ordinary life. In the modern liberal tradition, it is often considered a basic component of justice: a system of laws or constraints is just if, among other things, it is equitably applied to all concerned. Within the context of games however, fairness takes a different cast: that of ``fair play'', and ``sportsmanship'' or (in the gender-neutral synonym I'll use here) ``sportingness''. That players attempt to play the game in good faith is seen as an aspect of fairness and is frequently an implicit part of a game's ruleset. Under the logic of good-faith play, one may obey the explicit rules and still play unfairly, which helps distinguish this notion of fairness from the modern liberal tradition of fairness-as-justice. 

An example is provided by the bodyline bowling scandal of cricket in the 1930s~\cite{frith2013bodyline}. Bodyline bowling, as the name suggests, is the practice of aiming a bowled ball at the body of the opposing batsman, rather than the stumps. In doing so, the bowler forces the batsman to protect himself from injury, a natural defensive response that often leads to him being caught out by a nearby fieldsman. Bodyline bowling was not, at the time of its invention, forbidden by the rules of cricket, but was widely considered unfair and a violation of the good faith rule that the game should be decided on the basis of bat and ball, not physical intimidation through the threatened risk of injury.

In a related fashion, a soccer player who repeatedly and deliberately offsides and, in doing so, achieves an advantage over the opposing team, is unsporting even in the case that he does not attempt to deceive the referee and accepts the appropriate penalty as prescribed by the rules. If the violations are deliberate, rather than the occasional flaws of a good-faith player, then the reality of the game is put under stress. The most natural parallel in science to this kind of rule breaking are statistical tricks such as post-selection (dropping inconvenient results for contestable reasons, with foreknowledge that it will improve the apparent strength of the desired effect) and $p$-hacking.\footnote{Some bad behavior falls under a very different heading: not ``unsporting'', but cheating. Deliberately faking data by altering the contents of a spreadsheet is perhaps best seen by analogy to doping in physical competition, rather than pushing the boundaries of fair play.}

Nonetheless, the norms of sporting behavior and adherence to the assumption of good faith play also mean that we can expect players to be reluctant to draw attention to rule violations, particularly when they appear in a game episode that the player is not participating in. This is because by drawing attention to unsporting conduct she necessarily draws attention to the unreality of the game world and creates doubt about the very possibility of fair play. Rules are more than constraints: in being constitutive of the game, they have a further, quasi-sacred status that the players are required to uphold. A player who is observing a game from the outside cannot herself win the match by interfering, and interference risks destabilizing the conditions that would allow her to play, and possibly win, another match in the future. Calling out unsporting behavior is an extra-ludic act: neither part of the game's tension, nor its ruleset.

Such a disinclination to punish other players places games outside the usual domain of social activity. In ordinary transactions, non-participants play a crucial role in enforcing norms through both first- and second-order punishment~\cite{henrich2001people}---they punish cheaters even when they have not themselves been cheated and even judge others who refuse to punish---and this participation is frequently seen as necessary to the maintenance of community goods that they themselves enjoy (\emph{e.g.} law and order). In games, by contrast, people who consider themselves to share in the goods of a particular game community will avoid undertaking the task of punishment, because in damaging the necessary confidence in good-faith play, it may endanger the possibility of any win at all.

Taken together, a reliance on norms of sportingness and the extra-ludic nature of punishment predicts two key features of science as a form of play. We expect unsporting players to have bad reputations within their communities, but, at the same time, we also expect them not to receive public punishment by fellow players (in most circumstances). A scientist who is suspected of forging data may suffer indirectly though an absence of job offers and weak tenure letters, but will not be brought to trial by his colleagues. When cheaters \emph{are }subjected to scrutiny, it will often be from those outside the game community: very senior, retired members of the field, very junior members whose investment in the game is as yet relatively underdeveloped, or scientists whose ordinary domain of play is in a different field altogether. Finally, we expect unsporting conduct that \emph{is} publicly acknowledged to receive the most serious penalties: whatever the particulars of the transgression, all punishment shares the task of defending the logic of fair play.

The community's response to a scandal involving the scientist Brian Wansink provides an example of how unsporting conduct can lead to punishment. The questionable research practices (QRPs) engaged in by Wansink and his laboratory at Cornell included selective reporting of dependent variables, conditions, and results; post-hoc data exclusion; and reporting of unexpected correlations as being predicted from the start. Such behavior appears to be widespread within the psychology community~\cite{john2012measuring}, with self-reported rates of QRPs between 63\% for the ``most mild'' forms (failure to report all dependent variables) and 27\% for the more serious (post-hoc ``prediction'' of unexpected correlations). Wansink's practices were thus, at least at first glance, within the implicit norms of the community. His lab came under intense scrutiny in November of 2016~\cite{wansink}, however, after Wansink posted to his personal blog an article, ``The Grad Student Who Never Said ``No'''', describing these practices, and by the end of 2018 fifteen papers had been retracted.

From the standpoint of science as game play, the crime for which Wansink's lab was prosecuted was not actually the questionable research practices, but the explicit celebration of how these practices enabled him and his students to succeed. By punishing Wansink, the community prevented the emergence of common knowledge about the acceptability of unsporting research practices.

The game of sprinting the one-hundred meter dash provides an analogy to the Wansink case. The Olympic committee specifies that runners may not move from their blocks until 100 milliseconds after the gun has been fired; movement before that moment is considered to be physiologically impossible and thus to imply that the runner has (in an unsporting fashion) anticipated the gun. The continued existence of false starts indicates that runners still attempt to gain an edge in this fashion, however, and for many elite runners, their ``true'' reaction times may mean that a correctly-anticipated gun can gain them a winning position. Wansink is in the position of a running coach who boasts about training his athletes to become better at gun anticipation. Under this picture, it is how this common knowledge threatened the legitimacy of the game, rather than the epistemic problem behind the questionable practices, that led to his punishment. There are plenty of high-profile results that have influenced the field of psychology, are known within the community not to replicate, and yet are not retracted.

\section{Discussion}

Science is an epistemic phenomenon. Scientists are clearly involved in some sort of knowledge-formation process, and come to know things that were not known before. Science is also an economic phenomenon, in a broad sense: scientists recruit students to aid research, promote the approaches of their group, accumulate social capital, and both amass, and risk, power and influence.

Game play, however, is neither epistemic nor economic. It consists neither in the formation of beliefs nor the satisfaction of wants, though it may have both, or either, as a consequence. Among other things, this means that if scientists choose to play a game, they restrict themselves in ways that are invisible to a purely utility-theoretic account, and these constraints may help provide a more parsimonious description of what we see in practice.

Not all science is, thankfully, a game. One of the predictions of the Huizingan framework is that interdisciplinary science---which, at the very least, crosses pre-established game communities---is a way to prevent science from becoming a game, or to return it to a more open-ended process consistent with a Whewellian, Popperian, or Peircean vision. We might expect these fields to violate free engagement because of connections to the applied sciences, disconnection because of a concern to produce models that answer to an unbounded set of phenomena, boundedness because they share results outside the usual journal system, rule-constraint because they dynamically adjust criteria based on the results themselves, and tension because of a lack of consensus on what counts as publishable work. After some time, however, the establishment of interdisciplinary journals and the emergence of a new refereeing community around them may act to turn these anomalous practices into puzzle-solving games.

\section*{Acknowledgements}

I thank my two referees for helpful suggestions but not, honorably, for ``a game well played''. I acknowledge the support of the Templeton World Charity Foundation, and Jaan Tallinn via the Survival and Flourishing Fund.

\bibliographystyle{apacite}
%\bibliography{main-comments}

\begin{thebibliography}{}

\bibitem [\protect \citeauthoryear {%
Appiah%
}{%
Appiah%
}{%
{\protect \APACyear {2017}}%
}]{%
appiah2017if}
\APACinsertmetastar {%
appiah2017if}%
\begin{APACrefauthors}%
Appiah, K.%
\end{APACrefauthors}%
\unskip\
\newblock
\APACrefYear{2017}.
\newblock
\APACrefbtitle {As If: Idealization and Ideals} {As if: Idealization and
ideals}.
\newblock
\APACaddressPublisher{}{Harvard University Press}.
\PrintBackRefs{\CurrentBib}

\bibitem [\protect \citeauthoryear {%
Berezhiani%
\ \BBA {} Khoury%
}{%
Berezhiani%
\ \BBA {} Khoury%
}{%
{\protect \APACyear {2015}}%
}]{%
berezhiani2015theory}
\APACinsertmetastar {%
berezhiani2015theory}%
\begin{APACrefauthors}%
Berezhiani, L.%
\BCBT {}\ \BBA {} Khoury, J.%
\end{APACrefauthors}%
\unskip\
\newblock
\APACrefYearMonthDay{2015}{}{}.
\newblock
{\BBOQ}\APACrefatitle {Theory of dark matter superfluidity} {Theory of dark
matter superfluidity}.{\BBCQ}
\newblock
\APACjournalVolNumPages{Physical Review D}{92}{10}{103510}.
\PrintBackRefs{\CurrentBib}

\bibitem [\protect \citeauthoryear {%
Feyerabend%
}{%
Feyerabend%
}{%
{\protect \APACyear {1993}}%
}]{%
feyerabend1993against}
\APACinsertmetastar {%
feyerabend1993against}%
\begin{APACrefauthors}%
Feyerabend, P.%
\end{APACrefauthors}%
\unskip\
\newblock
\APACrefYear{1993}.
\newblock
\APACrefbtitle {Against method} {Against method}.
\newblock
\APACaddressPublisher{New York, NY, USA}{Verso}.
\PrintBackRefs{\CurrentBib}

\bibitem [\protect \citeauthoryear {%
Frith%
}{%
Frith%
}{%
{\protect \APACyear {2013}}%
}]{%
frith2013bodyline}
\APACinsertmetastar {%
frith2013bodyline}%
\begin{APACrefauthors}%
Frith, D.%
\end{APACrefauthors}%
\unskip\
\newblock
\APACrefYear{2013}.
\newblock
\APACrefbtitle {Bodyline Autopsy: The full story of the most sensational Test
cricket series: Australia v {E}ngland 1932-33} {Bodyline autopsy: The full
story of the most sensational test cricket series: Australia v {E}ngland
1932-33}.
\newblock
\APACaddressPublisher{New South Wales, Australia}{Aurum Press}.
\PrintBackRefs{\CurrentBib}

\bibitem [\protect \citeauthoryear {%
Geertz%
}{%
Geertz%
}{%
{\protect \APACyear {1972}}%
}]{%
geertz2005deep}
\APACinsertmetastar {%
geertz2005deep}%
\begin{APACrefauthors}%
Geertz, C.%
\end{APACrefauthors}%
\unskip\
\newblock
\APACrefYearMonthDay{1972}{}{}.
\newblock
{\BBOQ}\APACrefatitle {Deep play: Notes on the {B}alinese cockfight} {Deep
play: Notes on the {B}alinese cockfight}.{\BBCQ}
\newblock
\APACjournalVolNumPages{Daedalus}{101}{1}{1--37}.
\PrintBackRefs{\CurrentBib}

\bibitem [\protect \citeauthoryear {%
Gelman%
\ \BBA {} Shalizi%
}{%
Gelman%
\ \BBA {} Shalizi%
}{%
{\protect \APACyear {2013}}%
}]{%
gelman2013philosophy}
\APACinsertmetastar {%
gelman2013philosophy}%
\begin{APACrefauthors}%
Gelman, A.%
\BCBT {}\ \BBA {} Shalizi, C\BPBI R.%
\end{APACrefauthors}%
\unskip\
\newblock
\APACrefYearMonthDay{2013}{}{}.
\newblock
{\BBOQ}\APACrefatitle {Philosophy and the practice of {B}ayesian statistics}
{Philosophy and the practice of {B}ayesian statistics}.{\BBCQ}
\newblock
\APACjournalVolNumPages{British Journal of Mathematical and Statistical
Psychology}{66}{1}{8--38}.
\PrintBackRefs{\CurrentBib}

\bibitem [\protect \citeauthoryear {%
Gintis%
}{%
Gintis%
}{%
{\protect \APACyear {2000}}%
}]{%
gintis2000game}
\APACinsertmetastar {%
gintis2000game}%
\begin{APACrefauthors}%
Gintis, H.%
\end{APACrefauthors}%
\unskip\
\newblock
\APACrefYear{2000}.
\newblock
\APACrefbtitle {Game theory evolving: A problem-centered introduction to
modeling strategic behavior} {Game theory evolving: A problem-centered
introduction to modeling strategic behavior}.
\newblock
\APACaddressPublisher{Princeton, NJ, USA}{Princeton university press}.
\PrintBackRefs{\CurrentBib}

\bibitem [\protect \citeauthoryear {%
Gopnik%
}{%
Gopnik%
}{%
{\protect \APACyear {1996}}%
}]{%
gopnik1996scientist}
\APACinsertmetastar {%
gopnik1996scientist}%
\begin{APACrefauthors}%
Gopnik, A.%
\end{APACrefauthors}%
\unskip\
\newblock
\APACrefYearMonthDay{1996}{}{}.
\newblock
{\BBOQ}\APACrefatitle {The scientist as child} {The scientist as child}.{\BBCQ}
\newblock
\APACjournalVolNumPages{Philosophy of Science}{63}{4}{485--514}.
\PrintBackRefs{\CurrentBib}

\bibitem [\protect \citeauthoryear {%
Head%
, Holman%
, Lanfear%
, Kahn%
\BCBL {}\ \BBA {} Jennions%
}{%
Head%
\ \protect \BOthers {.}}{%
{\protect \APACyear {2015}}%
}]{%
phack}
\APACinsertmetastar {%
phack}%
\begin{APACrefauthors}%
Head, M\BPBI L.%
, Holman, L.%
, Lanfear, R.%
, Kahn, A\BPBI T.%
\BCBL {}\ \BBA {} Jennions, M\BPBI D.%
\end{APACrefauthors}%
\unskip\
\newblock
\APACrefYearMonthDay{2015}{03}{}.
\newblock
{\BBOQ}\APACrefatitle {The Extent and Consequences of P-Hacking in Science}
{The extent and consequences of p-hacking in science}.{\BBCQ}
\newblock
\APACjournalVolNumPages{PLOS Biology}{13}{3}{1-15}.
\newblock
\begin{APACrefDOI} \doi{10.1371/journal.pbio.1002106} \end{APACrefDOI}
\PrintBackRefs{\CurrentBib}

\bibitem [\protect \citeauthoryear {%
Henrich%
\ \BBA {} Boyd%
}{%
Henrich%
\ \BBA {} Boyd%
}{%
{\protect \APACyear {2001}}%
}]{%
henrich2001people}
\APACinsertmetastar {%
henrich2001people}%
\begin{APACrefauthors}%
Henrich, J.%
\BCBT {}\ \BBA {} Boyd, R.%
\end{APACrefauthors}%
\unskip\
\newblock
\APACrefYearMonthDay{2001}{}{}.
\newblock
{\BBOQ}\APACrefatitle {Why people punish defectors: Weak conformist
transmission can stabilize costly enforcement of norms in cooperative
dilemmas} {Why people punish defectors: Weak conformist transmission can
stabilize costly enforcement of norms in cooperative dilemmas}.{\BBCQ}
\newblock
\APACjournalVolNumPages{Journal of theoretical biology}{208}{1}{79--89}.
\PrintBackRefs{\CurrentBib}

\bibitem [\protect \citeauthoryear {%
Huizinga%
}{%
Huizinga%
}{%
{\protect \APACyear {2016}}%
}]{%
huizinga2016homo}
\APACinsertmetastar {%
huizinga2016homo}%
\begin{APACrefauthors}%
Huizinga, J.%
\end{APACrefauthors}%
\unskip\
\newblock
\APACrefYear{2016}.
\newblock
\APACrefbtitle {Homo Ludens: A Study of the Play-Element in Culture} {Homo
ludens: A study of the play-element in culture}.
\newblock
\APACaddressPublisher{Boston, MA, USA}{Beacon Press}.
\newblock
\APACrefnote{Translation from German edition, 1944}
\PrintBackRefs{\CurrentBib}

\bibitem [\protect \citeauthoryear {%
John%
, Loewenstein%
\BCBL {}\ \BBA {} Prelec%
}{%
John%
\ \protect \BOthers {.}}{%
{\protect \APACyear {2012}}%
}]{%
john2012measuring}
\APACinsertmetastar {%
john2012measuring}%
\begin{APACrefauthors}%
John, L\BPBI K.%
, Loewenstein, G.%
\BCBL {}\ \BBA {} Prelec, D.%
\end{APACrefauthors}%
\unskip\
\newblock
\APACrefYearMonthDay{2012}{}{}.
\newblock
{\BBOQ}\APACrefatitle {Measuring the prevalence of questionable research
practices with incentives for truth telling} {Measuring the prevalence of
questionable research practices with incentives for truth telling}.{\BBCQ}
\newblock
\APACjournalVolNumPages{Psychological science}{23}{5}{524--532}.
\PrintBackRefs{\CurrentBib}

\bibitem [\protect \citeauthoryear {%
Kauffman%
}{%
Kauffman%
}{%
{\protect \APACyear {2016}}%
}]{%
stu}
\APACinsertmetastar {%
stu}%
\begin{APACrefauthors}%
Kauffman, S\BPBI A.%
\end{APACrefauthors}%
\unskip\
\newblock
\APACrefYear{2016}.
\newblock
\APACrefbtitle {Humanity in a Creative Universe} {Humanity in a creative
universe}.
\newblock
\APACaddressPublisher{Oxford, UK}{Oxford University Press}.
\PrintBackRefs{\CurrentBib}

\bibitem [\protect \citeauthoryear {%
Kreps%
}{%
Kreps%
}{%
{\protect \APACyear {1990}}%
}]{%
kreps1990game}
\APACinsertmetastar {%
kreps1990game}%
\begin{APACrefauthors}%
Kreps, D.%
\end{APACrefauthors}%
\unskip\
\newblock
\APACrefYear{1990}.
\newblock
\APACrefbtitle {Game Theory and Economic Modelling} {Game theory and economic
modelling}.
\newblock
\APACaddressPublisher{Oxford, UK}{Clarendon Press}.
\PrintBackRefs{\CurrentBib}

\bibitem [\protect \citeauthoryear {%
Kuhn%
}{%
Kuhn%
}{%
{\protect \APACyear {2012}}%
}]{%
kuhn}
\APACinsertmetastar {%
kuhn}%
\begin{APACrefauthors}%
Kuhn, T.%
\end{APACrefauthors}%
\unskip\
\newblock
\APACrefYear{2012}.
\newblock
\APACrefbtitle {The Structure of Scientific Revolutions} {The structure of
scientific revolutions}.
\newblock
\APACaddressPublisher{Chicago, IL, USA}{University of Chicago Press}.
\newblock
\APACrefnote{Fourth edition. First published 1962}
\PrintBackRefs{\CurrentBib}

\bibitem [\protect \citeauthoryear {%
Maskin%
\ \BBA {} Tirole%
}{%
Maskin%
\ \BBA {} Tirole%
}{%
{\protect \APACyear {1988}}%
{\protect \APACexlab {{\protect \BCnt {1}}}}}]{%
maskin1988theory}
\APACinsertmetastar {%
maskin1988theory}%
\begin{APACrefauthors}%
Maskin, E.%
\BCBT {}\ \BBA {} Tirole, J.%
\end{APACrefauthors}%
\unskip\
\newblock
\APACrefYearMonthDay{1988{\protect \BCnt {1}}}{}{}.
\newblock
{\BBOQ}\APACrefatitle {A theory of dynamic oligopoly, {II}: Price competition,
kinked demand curves, and {E}dgeworth cycles} {A theory of dynamic oligopoly,
{II}: Price competition, kinked demand curves, and {E}dgeworth
cycles}.{\BBCQ}
\newblock
\APACjournalVolNumPages{Econometrica: Journal of the Econometric
Society}{}{}{571--599}.
\PrintBackRefs{\CurrentBib}

\bibitem [\protect \citeauthoryear {%
Maskin%
\ \BBA {} Tirole%
}{%
Maskin%
\ \BBA {} Tirole%
}{%
{\protect \APACyear {1988}}%
{\protect \APACexlab {{\protect \BCnt {2}}}}}]{%
maskin1988theory1}
\APACinsertmetastar {%
maskin1988theory1}%
\begin{APACrefauthors}%
Maskin, E.%
\BCBT {}\ \BBA {} Tirole, J.%
\end{APACrefauthors}%
\unskip\
\newblock
\APACrefYearMonthDay{1988{\protect \BCnt {2}}}{}{}.
\newblock
{\BBOQ}\APACrefatitle {A theory of dynamic oligopoly, {I}: Overview and
quantity competition with large fixed costs} {A theory of dynamic oligopoly,
{I}: Overview and quantity competition with large fixed costs}.{\BBCQ}
\newblock
\APACjournalVolNumPages{Econometrica: Journal of the Econometric
Society}{}{}{549--569}.
\PrintBackRefs{\CurrentBib}

\bibitem [\protect \citeauthoryear {%
McGaugh%
}{%
McGaugh%
}{%
{\protect \APACyear {2014}}%
}]{%
mcgaugh2014tale}
\APACinsertmetastar {%
mcgaugh2014tale}%
\begin{APACrefauthors}%
McGaugh, S\BPBI S.%
\end{APACrefauthors}%
\unskip\
\newblock
\APACrefYearMonthDay{2014}{}{}.
\newblock
{\BBOQ}\APACrefatitle {A tale of two paradigms: the mutual incommensurability
of {$\Lambda$CDM and MOND}} {A tale of two paradigms: the mutual
incommensurability of {$\Lambda$CDM and MOND}}.{\BBCQ}
\newblock
\APACjournalVolNumPages{Canadian Journal of Physics}{93}{2}{250--259}.
\PrintBackRefs{\CurrentBib}

\bibitem [\protect \citeauthoryear {%
Morgenstern%
\ \BBA {} Von~Neumann%
}{%
Morgenstern%
\ \BBA {} Von~Neumann%
}{%
{\protect \APACyear {1944}}%
}]{%
morgenstern2020theory}
\APACinsertmetastar {%
morgenstern2020theory}%
\begin{APACrefauthors}%
Morgenstern, O.%
\BCBT {}\ \BBA {} Von~Neumann, J.%
\end{APACrefauthors}%
\unskip\
\newblock
\APACrefYear{1944}.
\newblock
\APACrefbtitle {Theory of Games and Economic Behavior} {Theory of games and
economic behavior}.
\newblock
\APACaddressPublisher{}{Princeton University Press}.
\PrintBackRefs{\CurrentBib}

\bibitem [\protect \citeauthoryear {%
Peirce%
}{%
Peirce%
}{%
{\protect \APACyear {1868}}%
}]{%
peirce1868some}
\APACinsertmetastar {%
peirce1868some}%
\begin{APACrefauthors}%
Peirce, C\BPBI S.%
\end{APACrefauthors}%
\unskip\
\newblock
\APACrefYearMonthDay{1868}{}{}.
\newblock
{\BBOQ}\APACrefatitle {Some consequences of four incapacities} {Some
consequences of four incapacities}.{\BBCQ}
\newblock
\APACjournalVolNumPages{{The Journal of Speculative
Philosophy}}{2}{3}{140--157}.
\PrintBackRefs{\CurrentBib}

\bibitem [\protect \citeauthoryear {%
Peskin%
\ \BBA {} Schroeder%
}{%
Peskin%
\ \BBA {} Schroeder%
}{%
{\protect \APACyear {1995}}%
}]{%
peskin2018introduction}
\APACinsertmetastar {%
peskin2018introduction}%
\begin{APACrefauthors}%
Peskin, M.%
\BCBT {}\ \BBA {} Schroeder, D\BPBI V.%
\end{APACrefauthors}%
\unskip\
\newblock
\APACrefYear{1995}.
\newblock
\APACrefbtitle {An Introduction To Quantum Field Theory} {An introduction to
quantum field theory}.
\newblock
\APACaddressPublisher{New York, NY, USA}{CRC Press}.
\PrintBackRefs{\CurrentBib}

\bibitem [\protect \citeauthoryear {%
Ranka%
}{%
Ranka%
}{%
{\protect \APACyear {2008}}%
}]{%
ko}
\APACinsertmetastar {%
ko}%
\begin{APACrefauthors}%
Ranka.%
\end{APACrefauthors}%
\unskip\
\newblock
\APACrefYearMonthDay{2008}{}{}.
\newblock
{\BBOQ}\APACrefatitle {Interview with {Zhu Baoxun}} {Interview with {Zhu
Baoxun}}.{\BBCQ}
\newblock
\APACjournalVolNumPages{{RANKA} online: the Bulletin of the International Go
Federation}{}{}{}.
\newblock
\APACrefnote{\url{http://www.ranka.intergofed.org/?p=940}. Last accessed 3 June
2020}
\PrintBackRefs{\CurrentBib}

\bibitem [\protect \citeauthoryear {%
Roeckelein%
}{%
Roeckelein%
}{%
{\protect \APACyear {2006}}%
}]{%
roeckelein2006elsevier}
\APACinsertmetastar {%
roeckelein2006elsevier}%
\begin{APACrefauthors}%
Roeckelein, J.%
\end{APACrefauthors}%
\unskip\
\newblock
\APACrefYear{2006}.
\newblock
\APACrefbtitle {Elsevier's Dictionary of Psychological Theories} {Elsevier's
dictionary of psychological theories}.
\newblock
\APACaddressPublisher{}{Elsevier Science}.
\PrintBackRefs{\CurrentBib}

\bibitem [\protect \citeauthoryear {%
Smaldino%
\ \BBA {} McElreath%
}{%
Smaldino%
\ \BBA {} McElreath%
}{%
{\protect \APACyear {2016}}%
}]{%
smaldino2016natural}
\APACinsertmetastar {%
smaldino2016natural}%
\begin{APACrefauthors}%
Smaldino, P\BPBI E.%
\BCBT {}\ \BBA {} McElreath, R.%
\end{APACrefauthors}%
\unskip\
\newblock
\APACrefYearMonthDay{2016}{}{}.
\newblock
{\BBOQ}\APACrefatitle {The natural selection of bad science} {The natural
selection of bad science}.{\BBCQ}
\newblock
\APACjournalVolNumPages{Royal Society Open Science}{3}{9}{160384}.
\PrintBackRefs{\CurrentBib}

\bibitem [\protect \citeauthoryear {%
Wansink%
}{%
Wansink%
}{%
{\protect \APACyear {2016}}%
}]{%
wansink}
\APACinsertmetastar {%
wansink}%
\begin{APACrefauthors}%
Wansink, B.%
\end{APACrefauthors}%
\unskip\
\newblock
\APACrefYearMonthDay{2016}{}{}.
\newblock
\APACrefbtitle {The grad student who never said ``no''.} {The grad student who
never said ``no''.}
\newblock
\APACrefnote{Personal blog ``Healthier \& Happier''. Available at
\url{https://web.archive.org/web/20170312041524/http:/www.brianwansink.com/phd-advice/the-grad-student-who-never-said-no},
last accessed 12 Feburary 2019}
\PrintBackRefs{\CurrentBib}

\bibitem [\protect \citeauthoryear {%
Watson%
}{%
Watson%
}{%
{\protect \APACyear {2019}}%
}]{%
games_anthro}
\APACinsertmetastar {%
games_anthro}%
\begin{APACrefauthors}%
Watson, M.%
\end{APACrefauthors}%
\unskip\
\newblock
\APACrefYearMonthDay{2019}{}{}.
\newblock
{\BBOQ}\APACrefatitle {Games} {Games}.{\BBCQ}
\newblock
\BIn{} \APACrefbtitle {{The Cambridge Encyclopedia of Anthropology}.} {{The
Cambridge Encyclopedia of Anthropology}.}
\newblock
\APACaddressPublisher{}{University of Cambridge, Cambridge, UK}.
\newblock
\begin{APACrefDOI} \doi{10.29164/19games} \end{APACrefDOI}
\PrintBackRefs{\CurrentBib}

\bibitem [\protect \citeauthoryear {%
Whewell%
}{%
Whewell%
}{%
{\protect \APACyear {1847}}%
}]{%
whe}
\APACinsertmetastar {%
whe}%
\begin{APACrefauthors}%
Whewell, W.%
\end{APACrefauthors}%
\unskip\
\newblock
\APACrefYear{1847}.
\newblock
\APACrefbtitle {The Philosophy of the Inductive Sciences: Founded Upon Their
History} {The philosophy of the inductive sciences: Founded upon their
history}\ (\PrintOrdinal{2nd}\ \BEd, \BVOL~2).
\newblock
\APACaddressPublisher{London, United Kingdom}{John W. Parker, West Strand}.
\newblock
\APACrefnote{Page 65}
\PrintBackRefs{\CurrentBib}

\bibitem [\protect \citeauthoryear {%
Wojtowicz%
\ \BBA {} DeDeo%
}{%
Wojtowicz%
\ \BBA {} DeDeo%
}{%
{\protect \APACyear {2020}}%
}]{%
wojtowicz2020probability}
\APACinsertmetastar {%
wojtowicz2020probability}%
\begin{APACrefauthors}%
Wojtowicz, Z.%
\BCBT {}\ \BBA {} DeDeo, S.%
\end{APACrefauthors}%
\unskip\
\newblock
\APACrefYearMonthDay{2020}{}{}.
\newblock
{\BBOQ}\APACrefatitle {From Probability to Consilience: How Explanatory Values
Implement {B}ayesian Reasoning} {From probability to consilience: How
explanatory values implement {B}ayesian reasoning}.{\BBCQ}
\newblock
\APACjournalVolNumPages{{arXiv preprint}}{arXiv:2006.02359}{}{}.
\newblock
\APACrefnote{In review. \url{https://arxiv.org/abs/2006.02359}}
\PrintBackRefs{\CurrentBib}

\end{thebibliography}

%\section{}
%\subsection{}

\end{document}